\documentstyle [12pt]{article}
\evensidemargin\oddsidemargin
\begin{document} 
\count0 = 1    
\vspace{30mm}
 \title{{ Fuzzy Phase Space and Space-time Structure \\    
  as Approach to Quantization \\ }}
\author{S.N.Mayburov \\
Lebedev Inst. of Physics\\
Leninsky Prospect 53\\
Moscow, Russia, 117924\\
Email: mayburov@sci.lpi.msk.su
\\}
\date { }
\maketitle

\begin{abstract}

The quantum space-time
and the phase space of  massive particles  with Fuzzy Geometry
  structure  
investigated as the possible  quantization formalism.
 In this model the state of nonrelativistic
 particle $m$  corresponds to the element of
fuzzy ordered set (Foset)  - fuzzy point.
Due to its partial  ordering $m$ space coordinate 
$x$ acquires principal uncertainty $\sigma_x$.
It's shown that  Fuzzy Mechanics (FM) in 1+1 dimension 
 is equivalent to the path integral formalism of nonrelativistic 
Quantum Mechanics.  

\end{abstract}
\vspace{12mm}
\section { Introduction}

 The properties of quantum  space-time  and its  relation to
Quantum Mechanics (QM) axiomatics is
  actively discussed now from the different angles \cite {Vax,May,Aha}.
 The interest to it enforced
by the   indications that its structure  at small (Plank) scale
can be quite nontrivial \cite{Dop,Ish}.
In particular, it was proposed that
 such  fundamental 
 geometric features like the metrics or topology
 are modified significantly in this case    \cite {Ish,Vol}.
%
Our work motivated largely by this ideas and
 we shall explore the possible insights prompted by Sets Theory,
 exploring the various set structures  of space-time manifold $M$.
 For example, in 1-dimensional Euclidean Geometry, 
 the elements of its manifold $X$  - the points
$x_i$  constitute the  ordered set.
Yet Sets Theory includes  other kinds of sets  
which also permit to construct the  consistent geometries.
In this context we shall investigate  
  Posets and the fuzzy ordered sets (Fosets); in this case their elements 
are incomparable or weakly ordered relative to each other \cite {Zad,Got}.
Basing on Foset structure,  the
novel Fuzzy Geometry  was constructed   
which will be studied here as the possible space-time geometry
\cite {Zee,Dod,Pos}.


 In nonrelativistic Classical Mechanics in 1-dimensional space
 $X=R^1$ the particle's states corresponds to  the material  points $x^m(t)$
ordered relative to $R^1$  set, i.e relative to all its elements $\{x_a\}$.
In distinction on Fuzzy space manifold $M^F$ it supposed that
 the particle's  states correspond to the  fuzzy points $b_m$
 which   smeared 
in $R^1$ space  with an arbitrary dispersion $\sigma_x$.
In our approach the quantization regarded as the transfer from Classical
ordered phase space to fuzzy one.
In this paper as the  simple model of such transition
the  quantization of nonrelativistic particle will be regarded;  
 it will be argued that
Fuzzy Geometry  in 1-dimensional fuzzy space 
 induces the particle's dynamics which is equivalent to
Schr$\ddot{o}$dinger QM dynamics.
%
Earlier it was shown that  the  fuzzy observables 
are the natural generalization of QM observables \cite {Ali}.
The  'fuzzy lumps'  were applied in Quantum Gravity
and Cosmology studies \cite {Ren}. 
 In the last years it was shown
that some  fuzzy sets features
 are appropriate also for Quantum Logics formalism
 (\cite {Pyk} and ref. therein).

%

%

 Remind that in a partial ordered set (Poset)
$D=\{d_i\}$ beside the standard ordering relation between
its elements $d_k\leq d_l$ (or vice versa),
the  incomparability relation  $d_k\, \wr \,d_l$ is also permitted;
it means that both $d_k \le d_l$ and $d_l \le d_k$ propositions
 are false in this case. 
Fuzzy relations can be regarded as the generalization
of incomparability which
introduces  its continious measure $w$. To illustrate its meaning and
other Foset properties, let's
 consider the  discrete Poset $D^T$ which  includes  the 
 subset of incomparable elements $B=\{ b_j \} $,
and the   ordered subset $A=\{a_i \}$.
 Let's concede that in $A$  the elements    indexes grow
correspondingly to the elements order, so that for any $i$,  $a_i \le a_{i+1}$.
 Any $b_j\in B$ is incomparable at least to one $a_i \in A$.
The interval $[d_e,d_f]$ is $D^T$ subset, 
such that its maximal lower (upper) bound  $d_e, d_f \in A$,
and any $ a_i,\,b_j \in [d_e,d_f]$, if 
 $d_e\le a_i,\, b_j \le d_f$. For the simplicity let's consider $B$ which
 includes only  one element $b_0$. 
Let's suppose that   $b_0 \in [a_l,a_{l+n}];\,n \ge 2$,  $b_0$ 
is incomparable with all  $[a_{l},a_{l+n} ]$ internal elements:
  $ b_0 \wr a_i;\,$ iff $l+1\le i \le l+n-1 $,
 so that  $b_0$   in some sense is  'smeared' inside $[a_l,a_{l+n}]$
interval.

 To introduce  the  measure of incomparability $w$, let's put in correspondence
to each $b_0,a_i$ pair the weight $w^0_i \ge 0$ with the  norm $\sum w^0_i=1$.
The simplest  example is the symmetric incorability:  $w^0_i=\frac{1}{n}$ for 
 $a_i \in [a_l,a_{l+n}]$ interval;
$w^0_i=0$ outside of it; it can be interpreted as $b_0$ homogeneous smearing
inside $[a_l,a_{l+n}]$. If $b_0$ is  ordered (localized),
for example $b_0=a_i$, then $w^0_j=\delta_{ij}$.  
 If $w$ defined for all $a_j,b_i$ pairs in $D^T$, then
$D^T$ is  Foset, and $b_i$ are the fuzzy points.
 Note  that in distinction from 
the regarded case, in general an arbitrary Foset isn't necessarily Poset. 

The continuous 1-dimensional Foset $C^T$ can be defined
in the same vein;
the ordered  subset $A$ can be substituted
by the continuous metricized ( i.e. ordered) subset  $X \in C^T$ which is
equivalent to the real numbers axe $R^1$. In the simplest case one can
 take the same $B$, then $B \in C^T$ is the
 discrete subset of fuzzy points,  we shall regard here only
 $B=\{b_0\}$.  The interval $[x_v,x_u]$ is defined analogously to
 the discrete case, 
 the fuzzy $b_0,x_a$ relations are
described by the continuous distribution
 $w^0(x_a)\ge 0$ with the norm $\int w^0 dx=1$.
Below we shall call the fuzzy space $C^T = B \cup X$ which is
the direct sum of 1-dimensional Euclidean space and
the discrete set of  the fuzzy objects.
 For our study  the following example will be  important: let's
consider $b_0$ with $w^0$ support inside 
  the  system  of noncrossing intervals
 $E_x=\{ \bigcup \Delta_j ,\, j=1,...,n\}$,
then $b_0$ structure expressed by the relation  (proposition) :
$$
 LP^b:=
\quad b_0 \in E_x \,.and.\,
 b_0 \in \Delta_1.and.\,...\,.and. b_0 \in \Delta_n
$$
 so that $b_0$ can't be ascribed to any particular interval $\Delta_j$,
but only to their system $E_x$ as the whole.  
 %
Note that in Fuzzy Geometry  $w^0(x)$ doesn't have any probabilistic 
(stochastic) meaning but
only the topological and  geometric one. 
 The fuzzy   structures   to some extent are analogous
to Orthomodular or BvN algebras  which describes some Quantum 
Structures  \cite {Pyk}.
The remarkable analogy between the  uncertainty of 
Fuzzy points  coordinates and QM uncertainties
was noticed already \cite {Dod}, but no proof
of their equivalence was presented.
Fuzzy Geometry  formalism is the generalization
of regarded   examples and  is reviewed elsewhere \cite {Dod,Pos}. 
In brief form the main results of present paper were published in
\cite {May}


 \section {Fuzzy Mechanics (FM) and Fuzzy States}

Now we discuss the transition from Fuzzy geometry to Fuzzy mechanics (FM)
which is analogous to the transition from Euclidean Geometry to Classical Mechanics. 
The particle's state in Classical Mechanics is the ordered point
$\{\vec {r},\vec{p}\}$ in 6-dimensional phase space $R^3*R^3$.
We shall consider  the fuzzy point in coordinate space and the  
description of its evolution by some  state.
%
In nonrelativistic case considered here the time $t$ is the standard real
parameter on the axe $T$. We consider here  1-dimensional theory on $R^1$
and  suppose that  FM  posess the invariance relative to 
the space and time shifts and also is invariant
under space and time reflections
 analogously to Classical Mechanics; 
the particle evolution in FM is reversible \cite {Schiff}.  
%
In our theory  we identify the 
nonrelativistic particle $m$  with the fuzzy point  $b_0(t)$ in $C^T$
manifold described above;
it means that at any $t$ $m$ characterized by the positive
density $w(x,t)$  in 1-dimensional space $R^1$. But its evolution, as will
be shown below, can depends of more complicated correlations between
several $X$ points.
We concede that $m$ physical 
properties at the instant $t$  in an arbitrary  reference frame (RF)
  described by a fuzzy state $ |g (t)\}$, 
  the used notation stresses its difference from
Dirac quantum state $|\psi \rangle$.
The set of $|g\}$ states   $M_s$  doesn't supposed to be
 the linear space of any kind  $a\quad priory$, 
and one of our aims is to derive $M_s$ structure. In this approach
 $w(x,t)= F_N(g)$ is  some functional of $|g\}$,
 and  $|g\}$  have the positive constant norm: $N=\int wdx=1$.
Beside $w(x)$,  $g$ can include the additional
  components $\breve g$,  which will be used for the description
of $m$ evolution parameters; below  they will be related to 
  such $m$ observables as the  momentum $p$  and velocity $v_x$. 
We shall construct FM as the minimal theory i. e. at every stage
it assumed that the number of degrees of freedom and 
theory parameters is as minimal as necessary for the theory 
consistency. Consequently  we shall start  from assuming 
$\breve g$ to be the correlation field 
$\breve{g}=\{g_{\mu} (x_1,...,x_l)\};\, \mu=1,n$, where $g_{\mu}$ 
are some real functions, in the simplest case
 $\breve{g}=\{g_{\mu}(x)\}$ can be the vector field. 
  FM  formalism is  
based  mainly on geometric premises, analogously  to General Relativity. 
 In particular, the choice of  $\breve {g}$   
components will be motivated by Fuzzy Geometry. 
%

For the comparison of the observed effects,
besides the nonstochastic (pure) fuzzy states we shall consider also
the mixed fuzzy states $g^m$ which are
the probabilstic ensembles of several fuzzy states $\{g_i\}$
with probabilities $P_i$   and are analogous
to QM mixed states \cite {Schiff}. 
The measurement of $m$ observables in FM
   will be discussed in the final part of our
paper; here we assume only that ${x}$ 
 distributions  $w(x,t)$ can be measured by some experimental procedure.   
In addition, analogously to QM it supposed that an arbitrary $m$ state
can be prepared by some experimental procedure.

 
The evolution of any physical object can be described as the map
of its initial state $g_0$ to the final  $g(t)$;  so
the evolution of fuzzy point $m$ responds to the fuzzy map
$\Xi^f_t|g_0\} = |g(t)\}$. 
It's instructive to start from the study of  simple qualitative
 properties of such map.  
In particular, we consider the important effect of
   the sources smearing (SS) which is close analog of quantum  interference.
To illustrate  this effect which is generic for FM, let's study here 1-dimensional analog of the notorious  two slits experiment (TSE)
 which widely used  for  QM foundations discussion \cite {Schiff,Fey}.
 Here we regard   the initial state  
$|g_0\}$, which support  $E_x$ in some RF
 consists of $n_s$ noncrossing  intervals (bins)
$Dx_{i}$. 
After  $g_0$ preparation at $t=0$ presumably 
  $m$ doesn't interact with any other
object and evolves freely.
       $g_0$ can be regarded  as the source $S(g)$ for
the  future state -  the signal $|g(t)\}$, 
the resulting density $w(x,t)=\Xi^f_t g_0$.
 The fuzzy map $\Xi^f_t$ in principle can
    project   the internal fuzzy structure of the source
$S(g_0)$  to the distribution of signal density $w(x,t)$,
                     and due to it SS effect  will appear. 
For the simplicity we shall consider only an
infinitely small bins $Dx_{i} \rightarrow 0$,
so that $w^0_s$ can be approximated as: 
\begin {equation}
  w^0(x)=\sum \limits^{n_s} w^0_i\delta (x-x_i)  \label {D1} 
\end {equation}
Let's consider first  $n_s=1$, $w^0_1=1$ and suppose that in this case
FM evolution extends $g_0$ to $w(x,t)$  spread in some
 support $E_x(t)$, i.e. $w$ has the finite dispersion
$\sigma_x(t)$. Then it can be described as:
$$
       w_1(x,t)=C_m(\breve{g}^0,x-x_l,t)
$$
where  the function $C_m \ge 0$ is $w$ effective  propagator which
 conserves the norm:  
$$
       ||w_1||=\int C_m dx=1
$$ 
at any $t$.
Consider now the situation when $n_s \neq 1$ and
the signals from   different sources   don't intersect practically.
It's possible, for example, if $w_i$ dispersions $\sigma^i_x(t)$ 
are small, 
and  $|x_i-x_j| \to \infty$; then 
$$                     
   w_s(x,t) \simeq \sum w_i (x,t)
$$
From this properties,  in particular, from
$w$ norm conservation,   it's sensible to
assume that  $w_i(x,t)$ obeys to the relation:
$$
     w_i(x,t)=C_m(\bar{g}^0, x-x_i,t)w^0_i
$$
called here $w$-linearity.
Consider now $n_s=2$  case and suppose that 
 $w_{1,2}(x,t)$  for
  $m$ emission from $Dx_{1,2}$ at some $t$   intersect largely,
i.e.  $|x_1-x_2| \le \sigma_x^{1,2}(t)$. 
 What should be expected for the form of joint
distribution $w(x,t)$ in that case ?
If one prepares  the statistical mixture
of  $g^0_i$ states $|g^m_0\}$, in that case the weight $w^0_i=P_i$  
is the probability
for $m$ to be in $Dx_i$, and in each
individual event $m$  is  emitted definitely  by $Dx_1$ or $Dx_2$
 at $t_0$,  therefore for $g^m_0$ its  structure
is described by  the relation (proposition):
$$
LP^m := \quad m\in Dx_1 \, .or.\, m \in Dx_2
$$
Consequently, the final $m$ distribution
 will be the additive sum 
$$
 w^m(x,t)=w_1(x,t)+w_2(x,t)
=\sum w^0_i C_m(\bar{g}_i^0,x-x_i,t)
$$
 For the pure  initial $m$ state $g_0$ the following proposition
 describes the  source structure:
$$
 LP^p  := 
m \in D x_1 \, .and. \, m \in D x_2
$$
It follows that $LP^m$ and $LP^p$ are incompatible:
 $$
LP^p :\neq LP^m \,.or.\,LP^e
$$
for an arbitrary  proposition $LP^e$ which describes also some
$m$ signal.  The incompatibility of $LP^p,LP^m$ indicates that
the fuzzy source $S$ can't be decomposed into the sum of the local 
nonintersecting sources $Dx_{1,2}$; 
 the  density  $w_s(x,t)$  should have such form that it  makes
 in principle impossible 
to represent $w_s$ as the sum of two components which describes the signals
 from $Dx_{1,2}$  sources. It should be maximally
different  from the mixture $w^m$, so the $w^m$ content in $w_s$ 
 should be minimal, see lemma below.
 Therefore $w_s$
 should include the  nonlinear term $w_n$, for example, $w_n$
 can be proportional to $\sqrt{w^0_1w^0_2}$.
 feature of the fuzzy map. 

 Hence $g$ internal structure can be characterized by the parameter
$l^g=0,1$ for the mixed or pure states correspondingly. 
 In both cases  $w_s$ formulae for $n_s=2$ should be applicable
 for an arbitrary $w^0_i$,
 in particular, if  one of   $w^0_{i} \rightarrow 0$.
Therefore it decomposed as:
\begin {equation} 
   w_s(x,t)=w_1(x,t)+w_2(x,t)+l^g w_n(x,t)=w^m+l^gw_n
                                                         \label {EE2}
\end {equation}
where $w_n$ is FM nonlinear term. 
Due to $w_s$ norm conservation, the
resulting $w_n$ should obey the constraint $\int w_n dx=0$; 
it means that $w_n$  oscillates around $0$, and $w_s$ 
around $w^m$ correspondingly.

The importance of the  statements formulated above demands
 to prove them formally:\\
{\bf Lemma:} 
 $w_n(x,t)$ doesn't contain any linear combination
of $w_i(x,t)$; because of that $w^m$ content $k_m$ in $w_s$ is negligible.  
To prove the first proposition for $n_s=2$  let's suppose the opposite:
$$
   w_n(x,t)=w'_n(x,t)+\sum B^f_i(x-x_i,t)w^0_i
$$
     where $w'_n$ - the true  nonlinear component,
 $B^f_i(z,t) \ge 0$ are an arbitrary propagation functions. 
Then it means that :
$$
          w_s(x,t)= \sum (C_m+B_i)w^0_i+w'_n
$$
But this relation  should be true also for $w^0_i=\delta_{il};\, l=1,2$, 
corresponding effectively to $n_s=1$;
 in this case $w_s=w^m$, and from that $B_i=0$,
so if $w_n \ne 0$, then $w_n(x,t)$ is the nonlinear function of $w^0_i$. 
To demonstrate that $w_n \ne 0$ and $k_m=0$,
 let's rewrite $w_s$   in the form:
$$
w_s(x,t)=k_m w^m(x,t)+w_a(x,t)
$$
for $k_m\ge 0$ and an arbitrary $w_a(x,t) \ge 0$  which 
describes the signal from the source $S^a$ with an
arbitrary structure $LP^a$. 
The presence of   $w^m$ component in $w_s$
  implicates that: $LP^p := LP^m\,.or.\, LP^a$,
 but it contradicts with the obtained
  $LP^p, LP^m$ incompatibility.
To demonstrate that $|w_n| $ differs from $0$, let's consider
  its value inside $w_s$ support $E_s$, not including the points $x_m$ in 
which $w^m(x,t)=0$. If to assume that in the rest of $E_s$ $w_s(x) > 0$ 
then $w_s$ admits the solution with $k_m > 0$. To exclude it,
 it should exist at least one point $x_e$ for which $w_s(x_e)=0$, 
 from that follows $w_n(x_e) < 0 $. 
In other words the fuzzy map $\Xi^f_t$ permits 
any $w_s$ dependence on $w^0_i$ which conserve its norm,
 except the linear one, from
 which it should differ maximally.
It's natural to expect also that $w_n$ is local term in a sense that
$$
         w_n(x,t)=F[w_{1,2}(x,t)]
$$
notwithstanding the dependence on other $g$ components and
in our model we don't assume it.  However
   the weaker $w_n$ locality property, which also
follows from $w_n$ nonlinearity, will be used: if at some $x,t$
one of   $w_i(x,t) \to 0$, then   $w_n(x,t) \to 0$.
%

For illustration  FM evolution can be compared
with the classical diffusion \cite {Vlad}. Its  state is described by
$w(x,t)$ only, in this case for 
pointlike source in $x=0$ one obtains:
$$
     w^D(x,t)=\frac{1}{2\sqrt{\pi}k}\exp^{-\frac{x^2}{4k^2t}}
$$
where k is the diffusion constant. Naturally for this process 
for any initial 
$w_n=0$, yet $\sigma_x(t) \to \infty $ at $t \to \infty$.
One should define also SS measure i.e.
 the criteria of signals separation - $R_{ss}$ for the
evaluation of smearing rate; depending on it
$R_{ss}$ can vary  from $0$ to $1$. For $n_s=2$ it depends 
on the rate of $w_1,w_2$ overlap: 
$$
            R_w=\int \sqrt {w_1 w_2} dx
$$
in any realistic situation $R_{ss} \le R_w$, so to get the
maximal SS  it's necessary  that $R_w \rightarrow 1$ also.
The general $R_{ss}$ ansatz is quite complicated  \cite {Vax},  
but $R_{ss}$ will be used in our formalism  
only in the  asymptotic limits
 $R_{ss} \to 0$ or $1$,
 in that case one can take $R_{ss} \simeq R_w$.


From this considerations  we propose the simple  toy-model of FM which
helps to understand the main features of  FM evolution.
 Let's consider again  TSE  initial 
 state $g_0$ at $t_0$ with $w^0$ located in $n_s=2$ pointlike bins.
%
As was argued,   SS effects are generic for FM, and
 from that 
  FM free evolution  should be characterized by maximal SS,
Hence it's instructive to regard under which conditions 
it's possible in FM.  
Obviously for $n_s=2$ the maximal SS 
  for   an arbitrary, large
 $L_x=|x_1-x_2|$  is achieved, if $w_{i}$
dispersion $\sigma^i_x(t)$ is as large as possible
without violation of the theory consistency.
   
For example, if   $w_i= w^D(x,t)$ of
classical diffusion (but with resulting $w_n \ne 0$),
then this property is obvious.
In relativistic theory for an arbitrary localized $w_s$, the
 dispersion $\sigma^i_x (t)$ is restricted
by the maximal signals  velocity $c$, so that $\sigma^i_x \le ct$.
In  1-dimensioanl nonrelativistic theory 
 nothing forbids to choose  FM ansatz for $n_s=1$
with $\sigma_x(t) \rightarrow \infty$ at finite $t$. In this case
$w_{i}(x,t)$ should be Schwartz  distribution (generalized function) 
for which  at $x \rightarrow \pm \infty $,
 lim $w_i(x-x_i,t) \ne 0$ (or the limits don't exist) \cite {Schw}.
 This property is called   $x$-limit condition and the  class
of distributions which obeys it is denoted $W^x$.
In this case for $n_s=2$ $R_{ss}$ can be independent of $L_x$, even for
 $L_x \rightarrow \infty $,  so such theory doesn't need the additional
length  parameters.

To sharpen our arguments let's consider
 in this framework $n_s=1$ case, so that
 $w^s_0(x)=\delta(x-x_0)$. The resulting
 $w_s(x,t)$ is the distribution and obeys 
$x$-limit condition, therefore $\bar{x}(t)$ and higher $x$-moments
for it are undefined. 
Beside $x_0$, 
 $\Xi^f_t$ can project to $w(x,t)$  $g_0$ internal structure.
 Yet  as we assumed $g_0$ internal structure defined by its
geometry, then  for $n_s=1$
$g_0$ has the trivial geometric structure of the ordered point $x_0$,
 which means that it has no internal structure at all, in fact.
Therefore if only $x_0$ position defines $w_s(x,t)$ 
 and $w_s(x,t)\ne 0$ at
 $|x| \rightarrow \infty $,  then $w_s(x,t)$ should be a monotonously
 decreasing function of $\|x-x_0\|$.  
For example it can be $w_s \sim 1+c \exp(-\frac{(x-x_0)^2}{d_x^2})$;
yet as this example illustrates, any such continuous monotonous function
demands at least one length parameter $d_x$ ( which can also depend on $t$).
Yet the minimal FM geometrical  theory can be formulated 
without such fundamental length $d_x$, in that case the only solution    
 for such $g_0$  leads to $w_s(x,t)=const$ at any $t$.
It assume that FM can include the length scaling - i.e. conformal invariance
properties, and supports  $L_x \to \infty$ hypothesis.
Meanwhile $x_0$ value should be eventually mapped to $|g(t)\}$,
because for free FM evolution all
$g_0$ parameters should be extracted from $g(t)$.
Thus, $|g\}$ should contain at least one more degree of freedom $\breve{g}$
 beside $w(x)$, as was supposed above. 
 Such $g_0$ free evolution
at first sight seems quite exotic, remind yet that
 QM predicts the analogous evolution 
for the pointlike initial state \cite {Ber}.
Below this FM  model prediction for $n_s=1$ will not be used directly
in the formalism construction, but it will be obtained eventually
as the result of calculations.

Consider now  $n_s=2$ and the particular $x_{1,2},w^0_{1,2}$,
in our model $w_s(x,t)$ of (\ref {EE2}) obeys $x$-limit condition.
Let's choose the particular solution
 $w_s(x,t)$ which  responds to the maximal SS.
Then   $w'_s(x,t)=w_s(x+a_x,t)$ also responds to it  for an arbitrary $a_x$,
because
$R_{ss}$  depends of $w_s$ form only, not of $\bar{x}$.
Therefore $w'_s$ are also the solutions for $g_0$ evolution
problem, if it depends only on  $w_s^0$.
If $\sigma_x(t)$ is finite and $\bar{x}(t)$ is well defined,
 then it unambiguously stipulated by the initial state and its dynamics;
the example is the classical diffusion. 
But  $\bar{x}(t)$ and higher $x$-moments are undefined
 for $w_s$ which obeys $x$-limit condition, and 
in that case only $w_s$ form can depend on  
 FM dynamics;
 thereon $a_x$ value should be defined by the initial
$g_0$ components, but the alternative solution with an
arbitrary $a_x$ is consistent also.
This conclusion is especially obvious if $w_i(x,t)$ are
practically  independent of $x$, as our consideration of $n_s=1$
supposes. $w_s$ form can be characterized numerically by its
fourier-transform and other methods which details are unimportant here
 \cite {Ed}.
 It evidences that beside $w_0(x),\, g_0$   includes
  at least one more degree of freedom $\breve{g}$.
$a_x$ depend on $g$ both in $x_1$ and $x_2$, so if $\breve{g}$
responds to $a_x$ only, it is sensible to concede
that it can be represented as the binary  correlation $K^f(x_1,x_2)$.


 In the minimal theory  $K^f$ is an  arbitrary real
 function of two variables, which is  continuous or has
the finite number of breaking points.
Consequently,  for $n_s=2$ $a_x=f(K^f)$;
 if to choose the gauge: $\forall x_c;\,K^f(x_c,x_c)=0$, then 
 in 1-dimension one obtains for the  fixed $x_c$:
$$
  K^f(x_d,x_c) =\int\limits^{x_d}_{x_c}
 \frac{\partial K^f(\xi,x_c)}{\partial{\xi}} d\xi
$$
and from that:
$$
K^f(x_d,x_e)=K^f(x_d,x_c)-K^f(x_e,x_c)
$$
 Therefore  $K^f$ is, in fact,  the function of one observable:
$K^f(x,x_c) \rightarrow K(x)$. Because of it
 $g$ can be regarded as the local field
$E^g(x)=\{w(x),K(x)\}$.     
To simplify the evolution ansatz, we exploit
   the map $O E^g(x) \rightarrow g(x)$
 which parameters are defined below. It transforms $E^g$ into the 
 complex  function $ g(x)=g_1(x)+ig_2(x)$, where
 $g_{1,2}$ are the real functions.
$m$ density - $w(x)=F_w(g(x))$, where $F_w$ is  an  arbitrary
analytic  function.
If $g$ is the local field, it's natural  to assume that 
 $w/g$ zero-equivalence holds:
$w(x)=0 \,\Leftrightarrow \, g(x)=0$ and the same is true for $w,g$ limits
at $x \rightarrow \infty$.
In this case if  $x$-limit condition fulfilled for $w(x,t)$, then it's true
also for $g(x,t)$ which should be also Schwartz distribution.
 Of course, the alternative reasons for the additional $g$
degrees of freedom -  $K(x)$
appearance can exist, but in FM, as will be shown below,  
the proposed explanation is    the most appropriate one.
Note that even for the finite $\sigma_x(t)$ such $a_x$-dependent
ambigous solution can exist in the theory but their ansatz 
will be more complicated  \cite {Ed}.

Generally one should be careful with the interperetation of $w(x,t)$
distributions, as the measurable distributions of physical parameters,
 yet at this stage   it's admissable
to regard them at the same ground as
the standard, normalized distributions,
as was demonstrated in \cite {Fey}.
Below we shall reconsider this problem.
Regarded toy-model for $m$ free evolution
permit to assume that FM dynamics obeys the 'principle of
maximal fuzziness' (or minimal ordering)
 which can be formulated as the following:
at any $t$ $m$ state $g(t)$ characterized by the density $w(x,t)$ 
having the maximal $x$ uncertainty $\sigma_x(t)$ 
compatible with the initial $g_0$ structure.
Consequently, for $n_s=1$  $g_0$ has no internal structure
and induces $w(x,t)=const$  which has the maximal possible $x$
uncertainty.

\section {Particle Evolution in Fuzzy Mechanics}

After the semiqualitative FM toy-model consideration
we can turn to the fuzzy states $|g\}$  
 evolution formalism.
 From that discussion we concede that $m$ state
is described by a complex function $g(x)$ for which
$w/g$ zero-equivalence holds.  FM 
is assumed to be  invariant under $X,T$ shifts and  it  will be shown  that
 this  assumptions are enough 
 to calculate $m$ free evolution consitently.
We shall assume also that
 for $m$ free reversible evolution  $g(x,t)$ and its
Fourier-transform $\varphi(p,t)$ are continuous  for  
   $t \ge t_0$.
We'll argue that under this conditions  
 $g$ would  evolve in accordance with
 Schr$\ddot{o}$dinger free evolution operator
${U}_s(t)$ and $F_w(x,t)=|g(x,t)|^2 $. 
It would permit also to find the 
undefined $O$  map parameters.
In general   
 $g$ (or a state of any physical theory)
 free reversible evolution is described by the parameter-dependent operator
$U(t)$, so that: $g(t)=U(t)g_0$;
$U$, which correponds to $\Xi^f_t$,
 isn't supposed beforehand to be linear or unitary.
 However, it
possesses the properties of group elements: 
 $\forall t_{1,2};\, U(t_1+t_2)=U(t_1) U(t_2)$, therefore for
$m$ free evolution
$U(t)=e^{-i \hat{H}_0 t} $ where $\hat{H}_0$ is an arbitrary constant  
 operator \cite {Ber}. Meanwhile  the space shift operator $V$ is equal to:
 $V(a)=e^{a\frac{\partial}{ \partial x}}$,
for  the space shift of $g(x,t)$; from that $V(a)=e^{iap}$ when acting on
$\varphi(p,t)$ which is $g(x,t)$ Fourier-transform
\cite {Schiff}. The free evolution is invariant 
relative to those $X$ shifts,  because of it $U(t)$ commutes with 
$V(a)$ for the arbitrary $t,a$. It's equivalent to the
relation $[\hat{H}_0,p]=0$, from which follows that $\hat{H}_0$
for $\varphi(p,t)$ is an arbitrary function of $p$: $H_0=F_0(p)$.

 Consider now  the initial state $g_0$ for $n_s=1$: 
 the $w/g$ zero-equivalence at $t=t_0$ 
and the obvious condition $\varphi(p,t_0) \ne 0$ together
 result in  $g_0(x)=c_0\delta(x-x_0)$ for $w^0_s= \delta (x-x_0)$,
 where $c_0$ is an arbitrary complex
number (below we settle  $x_0=0, t_0=0$ for the simplicity).
Then, from well-known $\delta(x-x_0)$ Fourier transform
 $\varphi_{\delta}(p)=e^{ipx_0}$ it follows: 
 $$
       \varphi(p,t)=c_0 U(t)\varphi_{\delta}=c_0 e^{-iF_0(p)t}
$$
Let's study under which conditions the transition
 $ c_0\delta(x)\rightarrow g(x,t)$  develops smoothly and continiously.  
First, it  demands that $g(x,t_j)$  constitutes $\delta$-sequence, 
so that  $g(x,t_j)\rightarrow c_0\delta(x)$ for any
sequence   $\{t_j\} \rightarrow +0$ \cite {Ed}.
 It means that for an arbitrary function $\chi(x)$,
which belongs to the class of main functions  \cite {Vlad},
one has: 
$$
   I (\chi,t)=\int\limits^{\infty}_{-\infty}
 \chi(x) g(x,t) dx \rightarrow \chi(0)
$$
at $t \rightarrow +0$.
 It fulfilled only if $g(x,t)$ has $t=0$ pole,
so that $g(x,t)$ can be decomposed as: $g=g_s g_a$, where
$$
    g_s(x,t)=\frac{c_0} {f(t)} e^{i\gamma(z)} ,
$$
with an arbitrary, complex $\gamma$;
 $f(t)\rightarrow 0$
at $t \rightarrow 0$  so that for the substitution
$z=\frac{x}{f(t)}$ one has:
$$
\int\limits^{\infty}_{-\infty}
  g(z,t)f(t) dz \rightarrow 1
$$
at $t \rightarrow +0$ \cite {Vlad}.
 $g_a$ is an arbitrary, nonsingular function with
 $g_a[zf(t),t] \rightarrow  1$
  at $t \rightarrow +0$. Then under that conditions
$g(x,t) \rightarrow \delta(x)$ at $t \rightarrow +0$.
%
%
%
After $z$ substitution $g$ Fourier transform $\varphi$
alternatively can be  represented as:
\begin {eqnarray}  
\varphi'(p,t)= c_0 \int \limits^{\infty}_{-\infty} dz
g_a[zf(t),t] e^{i\gamma(z)+izpf(t)}= \exp^{-i\Gamma[pf(t),t]}
              \nonumber \\
\end {eqnarray}
Decomposing $g_a$ as the row in $t$, 
 from the equivalence  
  $\varphi (p,t) \simeq \varphi'(p,t)$  in the lowest $t$ order
one obtains  the equation:
\begin {eqnarray}
 \varphi(p,t)=e^{-i F_0(p) t} = \exp^{-i\Gamma[pf(t),0]}
 \nonumber \\
\end {eqnarray}
from which follows $F_0(p)=\frac {p^s}{2m_0}, f(t)=d_rt^r$
with $rs=1$, where $m_0,d_r$ are
an arbitrary  parameters. From that one finds $g_a(x,t)=1$, 
 $\Gamma(pf,t)=\Gamma(pf,0)$, and  the former equation
holds true at any $t$.
 If $H_0=F_0(p)$ is regarded as
 the free $m$ Hamiltonian,
then from its  symmetry properties
 the sensible $s$ values are only $s=2n$, where $n$ are the natural numbers
\cite {Schiff}.

%

Let's consider first the case $s=2$,
 it follows that the free Hamiltonian is  $H_0=\frac{ p^2}{2m_0}$ and    
  $U(t)$  is the unitary operator
for  the real $m_0$.
%
%
%
Therefore for $n_s=1$ and $g_0(x) = e^{i\alpha_0} \delta(x-x_0)$ one
obtains :
\begin {equation}
    g(x,t)= G(x-x_0,t)  e^{i\alpha_0} = \sqrt{\frac{m_0}{-i2\pi t}} 
 e^{i\frac{im_0(x-x_0)^2} {2t}+\alpha_0 }
             \label {DD23}
\end {equation}  
where for real, positive $m_0$ value the generalized function
 $G$ coincides with QM  free particle propagator \cite {Fey};
 it defines $\gamma,f$ completely.
Then in this formalism   $m_0$ can be interpreted as the  particle $m$ mass ;
note that for an imaginary $m_0$ such ansatz describes the classical diffusion.
%
 $g_0$ for the arbitrary $n_s$ can be written as :
$$
  g_0=\sum^{n_s} \sqrt{w_l^0}\delta(x-x_l)e^{i\alpha^0_l} 
$$
where $\alpha^0_l=K(x_l)$ are  an arbitrary real constants.
Obviously  one can transfer from the sum with
 $n_s\rightarrow \infty$ to  
  an arbitrary complex function for the initial state
 $g_0(x)=\sqrt{w_0(x)} e^{i\alpha^0(x)}$. In our formalism
 it evolves as: 
\begin {equation}
 g(x',t)= \int G(x'-x,t) g_0(x)dx
=\sqrt{ \frac{m_0}{-i2 \pi t}}
\int e^{\frac{im_0(x'-x)^2}{2t}} g_0(x)dx    \label {DD33}
\end {equation}
which coincides with the free $g_0$ evolution in QM path integral
formalism \cite {Fey}.
For such evolution ansatz one finds that the integral form
$N_2=\int |g(x,t)|^2dx$ is time independent and can settle $N_2=1$, meanwhile
in this case $F_w=|g|^2$ satisfies  to $m$ flow conservation
equation. Therefore $w(x)=F_w$
 can be chosen as $m$ universal density; in particular, it
permit to chose $c_0=1$.

Note that for $s=2, \, g(x,t) \neq 0$ at $x \rightarrow \pm \infty$,
i. e. satisfies to  $x$-limit condition, as our minimal FM assumes.
Yet it violated for free Hamiltonian with $s \geq 4$,
 in this case $g(x,t)$ asymptotics
can be  calculated \cite {Fed} at   $ x \to  \pm \infty$:
$$
   g(x,t) \simeq\frac{C_g}{ t^{\frac{1}{s}} }
(\frac{ t^{\frac{1}{s}} }{x} )^{\frac{s-2}{2(s-1)}}
\exp^{i\frac{s-1}{s} m^{\frac{1}{2(s-1)}}
 t^{-\frac{1}{2(s-1)}} {x}^\frac{s}{2(s-1)}}
$$
with $C_g$ - arbitrary constant.
In particular, for $s=4\,$, $|g| \sim \frac{1}{|x|^\frac{1}{3}}$.  
Therefore $g \to 0$ at $x \pm \infty$,   so it contradicts to $x$-limit
condition, and because of it the minimal FM assumptions are violated.
 In addition it can be shown that for $s \geq 4$
 it's impossible to construct $w(x)=F_w(g)$  as  a nonnegative,
 local $g$ form   which obeys the flow conservation. 
Therefore, all necessary conditions are fulfilled only for
$s=2$ (see also the calculations below).
  It turns out that  the obtained $U(t)$
 ansatz coincides with QM Schrodinger evolution
operator $U_s(t)$ for the free $m$ evolution.
Moreover, it agrees with the simple picture proposed in our FM toy-model.
 The analogous results for QM
are obtained  in the theory of the irreducible representations,
but  in that case they are based on more complicated axiomatics,
which includes axiom of RFs Galilean invariance \cite {Schiff}.
In distinction   Galilean Invariance for $g$ states in different RFs 
wasn't assumed in FM beforehand. It acknowledged in Quantum Physics
 that the classical
massiive objects, including physical RFs, can be regarded
as the quantum objects in the limit  $m \to \infty$ \cite {Schiff}.
If such approach is correct in FM framework also,
then regarding them as RFs, 
   Galilean transformations for them
  can be derived from the obtained FM ansatz for $H_0$.
Of course, this hypothesis needs further investigation,
 meanwhile in this approach it seems consistent.




Obtained $F_w(g)$ for $n_s=1$ gives $ w(x,t)  \sim \frac{1}{t}$ which
describes constant $m$ density analogously to QM results
for the initial state $g_0=\delta(x)$ \cite {Fey}.  
Therefore for $n_s=2$ 
$$
   w=\frac{1}{t}(w^0_1 + w^0_2 +2 \sqrt{w^0_1 w^0_2}
 \cos [ p^f_{12} (t) (x-\frac{x_{1}+x_2}{2})+ \alpha_{12}]
$$
where $p^f_{12}$ can be derived from (\ref {DD33}).
$w$  reproduces QM  sources interference;
it  corresponds  to the maximal SS  with $R_{ss}=1$ for $w^0_1=w^0_2$.
In physics  formally the distributions has the meaning 
only as  the functionals, so we should be careful with $w(x,t)$
interpretations. One can consider them  also as the limit of
the normalized observables  distributions; for example, $w(x)=const$
is the limit of Gaussian with $\sigma_x \rightarrow \infty$.
In this case $g(x,t)$ of (\ref {DD23}) as the state 
originating from from the narrow source can be substituted 
by the ansatz: 
$$
   g'(x',t)=  \int G(x'-x,t)e^{-\frac{x^2}{2\sigma_x^2}}dx 
$$
taken in the limit $\sigma_x \to 0$. Its main difference from 
$G(x,t)$ is that its norm $\int|g'(x',t)|^2dx'=1$.

Now  Hamilton formalism for FM can be formulated consistently. In our theory
from $X$ space shift symmetry it follows that
  $m$  momentum is the  operator
 $\hat{p}=i\frac{\partial}{\partial x}$ \cite {Schiff}
in $x$-representation
and  the free
 Hamiltonian  $\hat{H_0}=\frac{\hat{p}^2}{2m_0}$. 
 In FM the   natural $U(t)$  evolution generalization
 for the $m$ potential interactions $V_m(x)$ is:
 $\hat{H}=\hat{H}_0+V_m(x)$. From obtained relations it results in
Schr$\ddot{o}$dinger equation for $g$;
the general path integral ansatz for $g$ can be obtained
 by means of Langrangian $\cal L$  derived from $H$
 for the given  $V_e(x)$ \cite {Fey}.
The quantum phase $\alpha(x)$ properties acquires the natural description
in FM framework:
the real physical parameter is $K(x) \sim K^f(x,x')$ -
 the fuzzy correlation between 
 $x,x'$, and $\alpha=K(x)$ is its local  $x$
  representation  which is ambigous up to $2n\pi$.

Any complex normalized function $g(x)$ admits the orthogonal decomposition
on $|x_a\rangle=\delta(x-x_a)$, and $|x_a\rangle$ set constitute
the complete system \cite {Ber}. Therefore 
$|g\}$ set $M_s$   is equivalent to complex 
rigged Hilbert space $\cal{H}$  with  the scalar product
 $ g_1* g_2 = \int g_1^* g_2 dx$.
Consequently, our theory doesn't need Superposition Principle as
the independent axiom, its content follows from other FM axioms.
In FM  $x$   is $m$ observable and it's sensible
to suppose that $\hat p$ and any Hermitian operator function $\hat {Q}(x,p)$
 is also $m$ observable. For any such $Q$ there are exists
the correponding complete system of orthogonal eigenvectors $|q_a\rangle$
 in $\cal H$.
It permit to assume that for FM measurements description 
 QM reduction postulate for an arbitrary observable $Q$ 
can be incorporated in FM copiously,
so that  the 
particular outcome of $Q$ measurement $q_i$ 
appears with the probability $P=|\langle g| q_i\rangle|^2$  \cite {Schiff}.

%

  Generalization of
 FM formalism  on 3 dimensions is straightforward and will be regarded in the
 forcoming paper; here we  scatch only the main points.
We assume that the signal $g(x,t)$ from the point-like source 
 $w_0=\delta(\vec{r}-vec{r}_0)$ possess the spherical symmetry.
The correlation $\bar{g}$ between two points $\vec{r}_{1,2}$:
$$
  K^f(\vec{r}_1,\vec{r}_2) =\int\limits^{}_{l}
 \frac{\partial K^f(\vec{r},\vec{r}_2)}{\partial{\vec{r}}} d\vec{l}
$$
is supposed to be independent of the path $l$ over which it calculated. 
Analogously to 1-dimensional case $n_s=2$ we assume that for
the state from  two
pointlike sources 
 independently of the distance $|\vec{r}_1-\vec{r}_2|$ between them
achieved the maximal SS (calculated over $R^3$).

Note that Plank constant $\hbar=1$ in our FM calibration alike it's  
done in Relativistic QM; in any case it relates $x,p$ scales in the
regarded formalism \cite {Fey}.
 We believe  that the results obtained here can
have the general meaning for QM axiomatics considerations independently of
FM hypothesis.
The proposed FM considers the nonrelativistic particle for which 
$x$ is the fuzzy coordinate, yet from the symmetry of phase space
one can choose any observable $Q$ as the fundamental fuzzy
coordinate and from this assumption to reconstruct FM formalism.
It can be especially important in the relativistic case, where
, in distinction from $p$, $x$ can't the proper observable \cite {Schiff}.
 Such approach, in principle,
 can be extended on any physical system. In particular,
for the secondary quantization the numbers of 
certain particles $N_c(p)$ can be regarded as the fuzzy values.
  


\begin {thebibliography}{99}

\bibitem {Vax} S.Mayburov 'Fuzzy Space-Time Geometry
as Approach to Quantization',
 Proc. of Quantum Foundations conference,
( Vaxjio Univ. Press, Vaxjo, 2002), 233 - 242; hep-th 0210113 

\bibitem {May} S.Mayburov 'Fuzzy Geometry of Space-time and
Quantum Dynamics' Proc. Steklov Math. Inst. 245, 154 (2004)

\bibitem{Aha} Y.Aharonov, T.Kaufherr
'Reference Frames of Quantum Mechanics'
 Phys. Rev. D30 ,368 - 381 (1984)
 
\bibitem {Dop} S.Doplicher, K.Fredenhagen, K.Roberts
'Quantum  Space-Time at Plank Scale' 
 Comm. Math. Phys. 172,187-199 (1995) 


 

\bibitem {Ish} C.Isham
'Introduction into Canonnical Gravity Quantization'
 in: 'Canonical Gravity: from Classical to
Quantum ' Eds. J.Ehlers, H.Friedrich , Lecture Notes in Phys. 433, 11 - 27
(1994, Springer, Berlin)

\bibitem {Vol} V.S. Vladimirov, I.V. Volovich
'P-adic Quantum Mechanics', Comm. Math. Phys.
 123 - 132, 659 (1989) 


\bibitem {Zad} L.Zadeh 
'Fuzzy Sets'
Inform. and Control 8, 338 - 353 (1965)

\bibitem {Got} H.Bandemer, S.Gottwald 'Einfurlung in Fuzzy-Methoden'
(Academie Verlag, Berlin, 1993)


\bibitem {Zee} C.Zeeman,
'Topology of Brain and Visual Perception'
 in: ' Topology of 3-manifolds', 
ed. K.Fort, (Prentice-Hall, New Jersy, 1961), 240 - 256

\bibitem {Dod} C.T.J. Dodson,
'Tangent Structures for Hazy Space'
 J. London Math. Soc., 2,465 - 472 (1975)

\bibitem {Pos} T.Poston 'Fuzzy Geometry', Manifold, 10, 25 (1971)

\bibitem {Ali} T.Ali, G. Emch
'Fuzzy Observables in Quantum Mechanics'
 Journ. Math. Phys. 15, 176 - 182 (1974)
\, ibid., 18, 219, (1977)

\bibitem {Pyk} J.Pykaz,
'Lukasiewics Operations on Fuzzy Sets'
 Found. Phys. 30, 1503 - 1519 (2000)

\bibitem {Ren} M.Requardt , S.Roy
'Quantum Space-Time as Statistical Geometry of Fuzzy Lumps'
  Class. Quant. Grav  18 , 3039 -3058 (2001)




\bibitem {Schiff} J.M.Jauch, 'Foundations of Quantum Mechanics'
 (Adison-Wesly, Reading, 1968) 


\bibitem {Vlad} V.S.Vladimirov 'Equation of Mathematical Physics'
(Nauka, Moscow, 1971)

\bibitem {Ber} F.Berezin, E.Shubin
 'Schr$\ddot{o}$dinger Equation' (Moscow, Nauka, 1985) 

\bibitem {Schw} L.Schwartz 'Methods Mathematique pour
 les Sciences Physique' (Hermann, Paris, 1961)


\bibitem {Fed} M.Fedoriuk, 'Asymptotics: Intgrals and raws'
(Nauka, Moscow, 1987)

 \bibitem {Ed} R.Edwards 'Functional Analysis and Applications'
 (N-Y, McGrow-Hill, 1965)

\bibitem {Fey} R.Feynman,A.Hibbs 'Quantum Mechanics and Path Integrals'
(N-y, Mcgrow-Hill,1965)




 

\end {thebibliography}
\end {document}